\newcommand{\dis}{\displaystyle}
\newcommand{\mc}{\cal}
\documentstyle[12pt,epsf]{article}

\begin{document}
\parindent 1.3cm
\thispagestyle{empty}   
\vspace*{-3cm}
\noindent

\def\arccot{\mathop{\rm arccot}\nolimits}
\def\sd{\strut\displaystyle}

\begin{obeylines}
\begin{flushright}
UG-FT-83/97 \\
hep-ph/9712275 \\
November 1997
\end{flushright}
\end{obeylines}
\vspace{3.5cm}

\begin{center} \begin{bf} \noindent PHYSICS BEYOND THE STANDARD MODEL AT HERA
\footnote{Talk presented by F. Cornet at the XXI International School on
Theoretical Physics, Ustro\'n (Poland), September 1997.} \end{bf}
\vspace{1.5cm}\\ FERNANDO CORNET and JAVIER RICO \footnote{E-mail addresses:
cornet@ugr.es; jrico@rigoberto.ugr.es} \vspace{0.1cm}\\ Departamento de F\'\i
sica Te\'orica y del Cosmos,\\ Universidad de Granada, 18071 Granada, Spain\\
\vspace{2.2cm}

{\bf ABSTRACT}

\parbox[t]{12cm}{
This talk is divided in two parts. In the first one
we discuss the signals of the Minimal Supersymmetric Standard
Model through the production of $\tilde{e} \tilde{q}$. The second
part is devoted to contact terms. The bounds on the
mass scale $\Lambda$ obtained from atomic parity violation experiments and from
LEP are reviewed. Afterwards, we show that the excess of
events at high $Q^2$ observed at HERA could be explained in
terms of these contact terms.}
\end{center}

\newpage
\section{Introduction}
The potential of HERA to discover some hints of Physics Beyond
the Standard Model has been an important subject in all the
studies and workshops on the HERA physics potential. In a single talk we
cannot cover all the work that has been done in all these years. Rather,
we refer to the interested readers to an early report on the subject
\cite{CASHMORE} and to the proceedings of the workshops
on HERA physics \cite{1987,1991,1996} as well as to the references therein.

HERA is an $ep$ collider with a $27.5 \; GeV$ electron (or positron) beam and
an $820 \; GeV$ proton beam that gives a center of mass energy of
$300 \; GeV$. This energy allows to probe the proton down to distances
$O(10^{-16}) \; cm$.
The total luminosities collected up to now by each experiment
are $23.7 \; pb^{-1}$ (H1) and $33.5 \; pb^{-1}$ (ZEUS). An upgrade in
luminosity is planned for the year 2000. The hope is to reach a yearly
integrated luminosity of order $150 \; pb^{-1}$. This, obviously, would improve
very much the potential to discover signals of new physics.

The peculiar initial state at the partonic level, $eq$ or $eg$, forces most
of the effects of new physics to appear through a $t$-channel exchange of
new particles (this is the case for new gauge bosons \cite{RUCKL}) or
in associate production. The only particles that can be produced in the
$s$-channel are leptoquarks or leptoquark-like, i.e. particles that
couple to a quark and a lepton (such as squarks in R-parity violating
SUSY).

Recently, both experiments at HERA, H1 and ZEUS, have announced the observation
of an excess of events with respect to the Standard Model prediction
at very high $Q^2$ \cite{H1,ZEUS,CONF} in $e^+p$
neutral current and charged current deep inelastic scattering. Each experiment
observe 18 $e^+ p \to e^+ X$ events with $Q^2 > 15000 \; GeV^2$,
while only 8 are expected at H1 and 15 at ZEUS. In charged current deep
inelastic scattering, H1 observes 6 events with $Q^2 > 15000 \; GeV^2$
and expects 3, while ZEUS observes 5 events with the same $Q^2_{min}$ and
expects 2.
We will not elaborate any longer on the kinematics of these events because
an updated description of the experimental situation can be found in
Hubert Spiesberger's talk in this school \cite{SPIESBERGER}. In this talk you
can also find a discussion on the attempts to explain the excess of events
within the standard model, modifying the proton structure functions
\cite{SM}. However, none of the proposed modifications produces an increase
in the cross-section large enough to explain the data. Thus, if we assume
that the large number of events observed is not a statistical fluctuation,
we have to invoke some new physics to explain it. Three main ideas have
been proposed: the $s$-channel production of a leptoquark \cite{LEPTOQUARK},
the $s$-channel production of an squark within $R$-parity violating SUSY
\cite{SQUARK} and the presence of a non-resonant, four-fermion contact
interaction \cite{CONTACT}. We refer to the reader to the talks of
Hubert Spiesberger \cite{SPIESBERGER} and Jan Kalinowski \cite{KALINOWSKI}
for a detailed discussion of the first
two possibilities. In the third section of this talk we will present the
contact term interpretation.

The structure of this talk is the following. After this introduction we will
present a summary of the work that were performed in the last HERA workshop
studying the signals you can expect from the Minimal Supersymmetric
Standard Model (MSSM), with $R$-parity conservation. The third section
will be devoted to contact interactions, both in neutral current and charged
current deep inelastic scattering. Finally, we will finish with a brief summary
\footnote{The signals of a non-standard light Higgs boson at HERA has been
presented by M. Krawczyk at this school \cite{KRAWCZYK}.}.

\section{Minimal Supersymmetric Standard Model}
Supersymmetry (SUSY) is one of the main candidates for physics beyond the
Standard Model. A very clear signature of the new fermion-boson
symmetry is the doubling of particles. Each particle in the Standard Model
is predicted to have, at least, a superpartner which differs in half a unit
of spin from the standard particle.
The discovery of these superpartners would provide an
unequivocal proof of SUSY. There has been intensive search for them but,
up to now, no superpartner has been found. HERA should contribute in this
search and there has been some studies on the sensitivity that can be expected
for different processes. In this talk we will concentrate on the Minimal
Supersymmetric Standard Model with $R$-parity conservation.

The most promising process in $ep$ collisions is associate production
of selectron and squark \cite{STANCO,BARTL,EISENBERG,SCHLEPER}
\begin{equation}
\label{PROC}
e p \to \tilde{e} \tilde{q} X.
\end{equation}
It proceeds through neutralino $t$-channel exchange. Normally, the exchange of
the lightest neutralino would give the largest contribution, unless it turns
out
to be a higgsino. In this case the smallness of the higgsino couplings to the
electron and the light quarks in the proton suppresses its contribution in
front of the exchange of heavier neutralinos with larger couplings. The net
effect is that in the case that the lightest neutralino is dominated by a
higgsino, the cross-section for the process (\ref{PROC}) is too small to
be measured at HERA.

The masses and couplings of the neutralinos depend on the parameters:
$M_1$, $M_2$, $\mu$ and $ \tan \beta = \dis{v_2 \over v_1}$, that are the
$U(1)$ and $SU(2)$ gaugino masses, the higgsino mass parameter and the ratio
between the vacuum expectation values of the two Higgs doublets, respectively.
The two gaugino masses are related in Grand Unified Theories through:
\begin{equation}
\label{GUT}
M_1 = \dis{5 \over 3} \tan^2 \theta_W M_2
\end{equation}

The cross-section for the process (\ref{PROC}) depends on the sum of the
selectron and squark masses and on the gaugino mass parameter, say $M_2$,
but depends very weakly on each independent sfermion mass and $\tan \beta$.
Moreover, since the higgsino contribution is suppressed,
$\sigma (e p \to \tilde{e} \tilde{q} X)$ depends also very weakly on
$\mu$ in the accessible region of the parameter space.

The produced sfermions decay into the corresponding fermions and a neutralino,
or another fermion and a chargino:
\begin{equation}
\label{DECAYS}
\begin{array}{ll}
\tilde{e} \to e \chi^0_i \quad i = 1, \dots 4 &
\tilde{e} \to \nu \chi^\pm_i \quad i = 1,2 \\[0.3cm]
\tilde{q} \to q \chi^0_i \quad i = 1, \dots 4 &
\tilde{q} \to q^\prime \chi^\pm_i \quad i = 1,2 .
\end{array}
\end{equation}
We will assume that the lightest neutralino ($\chi^0_1$) is the lightest
SUSY particle. Since $R$-parity is assumed to be conserved $\chi^0_1$ is,
thus, a stable particle that once it is produced will escape the
detector as missing energy. The decay channel into the lightest neutralino
provides the clearest signature for selectron squark production. The optimal
situation to search for the process (\ref{PROC}) is that only $\chi^0_1$
is lighter than the sfermions and the other neutralinos and charginos are
heavier than the sfermions. If this is not the case and decays into another
neutralinos and charginos are kinematically allowed, these $\chi^0_i$ with
$i>1$ and $\chi^\pm_i$ would chain-decay into fermions and $\chi^0_1$. The
final state, thus, would contain a large number of fermions and missing
energy. Since this final state is much more difficult to search for than the
simple decay channel $\tilde{f} \to f \chi^0_1$, the analysis performed
in Ref.\ \cite{SCHLEPER} has been done searching only for this latter decay
channel. The contribution from the other channels has been taken into account
via the corresponding decrease of the branching ratio into $f \chi^0_1$. The
value of this branching ratio can again be calculated in terms of the same
parameters discussed above.

A systematic survey of the parameter space was performed in
Ref.\ \cite{SCHLEPER} allowing the parameters $M_{\tilde{e}}$,
$M_{\tilde{q}}$ and $M_2$ to vary within the bounds
$45 \le M_{\tilde{e}} , M_{\tilde{q}} \le 160 \; GeV$ and
$0 \le M_2 \le 200 \; GeV$ for the extreme values of $\mu =- 300 \; GeV$,
and $ 300 \; GeV$, assuming that the partners of the right and
left-handed fermions have the same mass. The values of $\tan \beta$ were
taken to be $\tan \beta = 1.41$ and $35$ \footnote{Some experimental
bounds have already been obtained for this process by H1 \cite{H1SUSY}
and ZEUS \cite{ZEUSSUSY}.
However, the total integrated luminosity collected up to now is too small
to allow for competitive bounds.}. The exclusion region in the plane
$(M_{\tilde{e}}+M_{\tilde{q}})/2 , M_2$ are shown in Fig.\ 1 for
$\tan \beta = 1.41$ and $\mu = -300 \; GeV$ ($\chi^0_1$ is photino-like).
The smallest triangle is the region already excluded by LEP1 and two
integrated luminosities have been assumed: $100 \; pb^{-1}$ and
$500 \; pb^{-1}$. The diagonal line is the limit where $\chi_1^0$ has
the same mass as the selectron or squark. The kink you can see for
small $M_2$ values is due to loss in sensitivity caused  by the opening
of sfermion decay channels other that $\tilde{f} \to f \chi^0_1$.

\begin{figure}
\setlength{\unitlength}{1cm}
\begin{picture}(12.5,8.)
\epsfxsize=13.5cm
\put(1.64,-8.8){\epsfbox{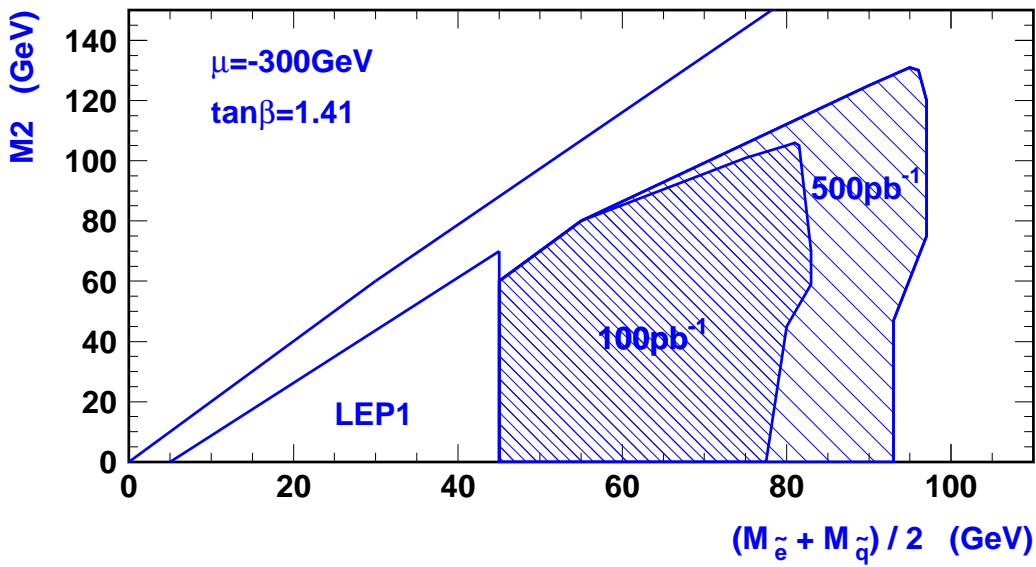}}
\end{picture}
\caption{Exclusion limits for HERA from the process
$ep \to \tilde{e} \tilde{q}X$ at $95 \% C.L.$ assuming two integrated
luminosities $100 \; pb^{-1}$ and $500 \; pb^{-1}$. The small triangle
is the region already excluded by LEP1 and the diagonal line is the limit
where the lightest neutralino has the same mass as the selectron or squark.
Figure taken from Ref.\ \cite{SCHLEPER}.}
\end{figure}

We have already stated that the cross-section depends very weakly on
$\mu$ and $\tan \beta$. Thus, the bounds shown in Fig.\ 1 also depend
weakly on those parameters. For instance, changing $\tan \beta$ from
$1.41$ (shown in the  figure) to $\tan \beta = 35$ the bound on $M_2$
only gets around $10 \; GeV$ higher and changing $\mu = -300 \; GeV$ to
$\mu = 300 \; GeV$ (i.e. going from a photino-like to $\tilde{Z}$-like
lightest neutralino), the bounds on $(M_{\tilde{e}}+M_{\tilde{q}})/2$
are also around $10 \; GeV$ higher. Finally, relaxing the GUT relation
(\ref{GUT}) does not change the results as long as the mass of the
lightest neutralino (which is the parameter that really enters in the
cross-section calculation) is not changed.

The bounds shown in Fig.\ 1 for $M_2$  are more stringent than
the ones that can be obtained at LEP2 in chargino searches. However,
if $\chi^0_1$ is $\tilde{Z}$-like ($\mu = 300 \; GeV$)
the bound that can be obtained at
LEP2 becomes stronger than the ones discussed here. The sensitivity
of HERA to selectron and squark masses turns out to be very similar
to the one of LEP2. However, it is interesting to note that the
production mechanisms are completely different. In such a way that
if a signal is found measurements at both colliders complement in the
study of the Minimal Supersymmetric Standard Model parameters.
The squark production cross-section at TEVATRON is very large. However,
in order to have a clear signal one has to cut in the mass of
$\chi^0_1$. At HERA, instead, you can be sensitive to larger values
of this mass.

\section{Contact Terms}
The low energy effects of physics beyond the SM, characterized by a
mass scale $\Lambda$ much larger than the Fermi Scale can be parametrized
in a very general way in terms of a non-renormalizable, effective lagrangian.
In this lagrangian all the operators are organized according to their
dimensionality in such a way that the higher dimension operators are suppressed
by powers of $\dis{{1 \over \Lambda}}$. Since the momenta involved in present
experiments are much lower than $\Lambda$ one expects the lower dimension
operators to give the bulk of the new physics effects. The relevant
lagrangian for
$ep$ scattering, including dimension $6$, four-fermion operators is
\footnote{Terms with scalar currents can also contribute to this order,
but we will not consider them here.}:
\begin{equation}
{\mc L} = {\mc L}_{SM} + {\mc L}_V ,
\end{equation}
where ${\mc L}_{SM}$ is the SM lagrangian and ${\mc L}_V$ is given by:
\begin{equation}
\label{LAG}
\begin{array}{lcl}
{\mc L}_V & = & \eta_{LL}^{(3),q} (\bar{l}\gamma_{\mu}\tau^{I}l)
               (\bar{q}\gamma^{\mu}\tau^{I}q) +
	       \eta_{LL}^{(1),q}(\bar{l}\gamma_{\mu}l)(\bar{q}\gamma^{\mu}q) \\
         & + & \eta_{LR}^{u}(\bar{l}\gamma_{\mu}l)(\bar{u}\gamma^{\mu}u) +
	       \eta_{LR}^{d}(\bar{l}\gamma_{\mu}l)(\bar{d}\gamma^{\mu}d) \\
         & + & \eta_{RL}^{q}(\bar{e}\gamma_{\mu}e)(\bar{q}\gamma^{\mu}q) \\
         & + & \eta_{RR}^{u}(\bar{e}\gamma_{\mu}e)(\bar{u}\gamma^{\mu}u) +
               \eta_{RR}^{d}(\bar{e}\gamma_{\mu}e)(\bar{d}\gamma^{\mu}d)  .
\end{array}
\end{equation}
\noindent
The $SU(2)$ doublets $l=(\nu,e)$
and $q=(u,d)$ denote left-handed
fields ($Ll = l, Lq = q, \hbox{with}\ L=\frac{1-\gamma_5}{2}$) and the SU(2)
singlets $e$, $u$ and $d$ represent right-handed electron, up-quark and
down-quark ($Re=e, Ru=u, Rd=d, \hbox{with}\ R = \frac{1+\gamma_5}{2}$).
It is customary to replace the coefficients $\eta$ by a mass scale $\Lambda$:
\begin{equation}
\eta = {\epsilon g^2 \over \Lambda^2},
\end{equation}
with $\epsilon = \pm 1$ taking into account the two possible interference
patterns. $\Lambda$ is interpreted as the mass scale for new physics in the
strong coupling regime, i.e.\ with
\begin{equation}
{g^2 \over 4 \pi} = 1.
\end{equation}
The lagrangian in Eq.\ (\ref{LAG}) modifies both, the neutral current and the
charged current cross-sections. We will now discuss both cases.

\subsection{Neutral Currents}
Contact terms have been proposed as a possible explanation for the excess
of events at high $Q^2$ observed at HERA \cite{CONTACT}. With the new data
the first indications towards a resonance structure in the events seem
to disappear, thus giving more strength to the contact term hypothesis.
The cross-sections for the processes $e^- p \to e^- X$ and $e^+ p \to e^+ X$
receive contributions from all the terms in the $O(p^6)$ lagrangian. However,
there is an important difference between both  cross sections.
The contributions from the $LR$ and $RL$ contact terms are suppressed in
the $e^-$ cross-section, while they give the largest contribution to the
$e^+$ cross-section \cite{RUCKL1,CORNET1,MARTYN}. This can be explicitly
seen in Fig.\ 2 where we show
\begin{equation}
\label{F}
F(x,Q^2) = \dis{ s x^2 y^2 \dis{d \sigma \over dx dQ^2} \over
                 2 \pi \alpha^2 (1+(1-y)^2)}
\end{equation}
for $x=0.3$ as a function of $Q^2$ for the six models we are considering with
$ \epsilon =1$ and $\Lambda = 3 \; TeV$. The variable $y$ is defined as usual:
$y = \dis{Q^2 \over sx}$. Although, we have shown in the figure only a value
of $x$ this is true in all the interesting values of $Q^2$ and $x$.
\begin{figure}
\setlength{\unitlength}{1cm}
\begin{picture}(12.,19.)
\epsfxsize=14cm
\put(1.2,0.){\epsfbox{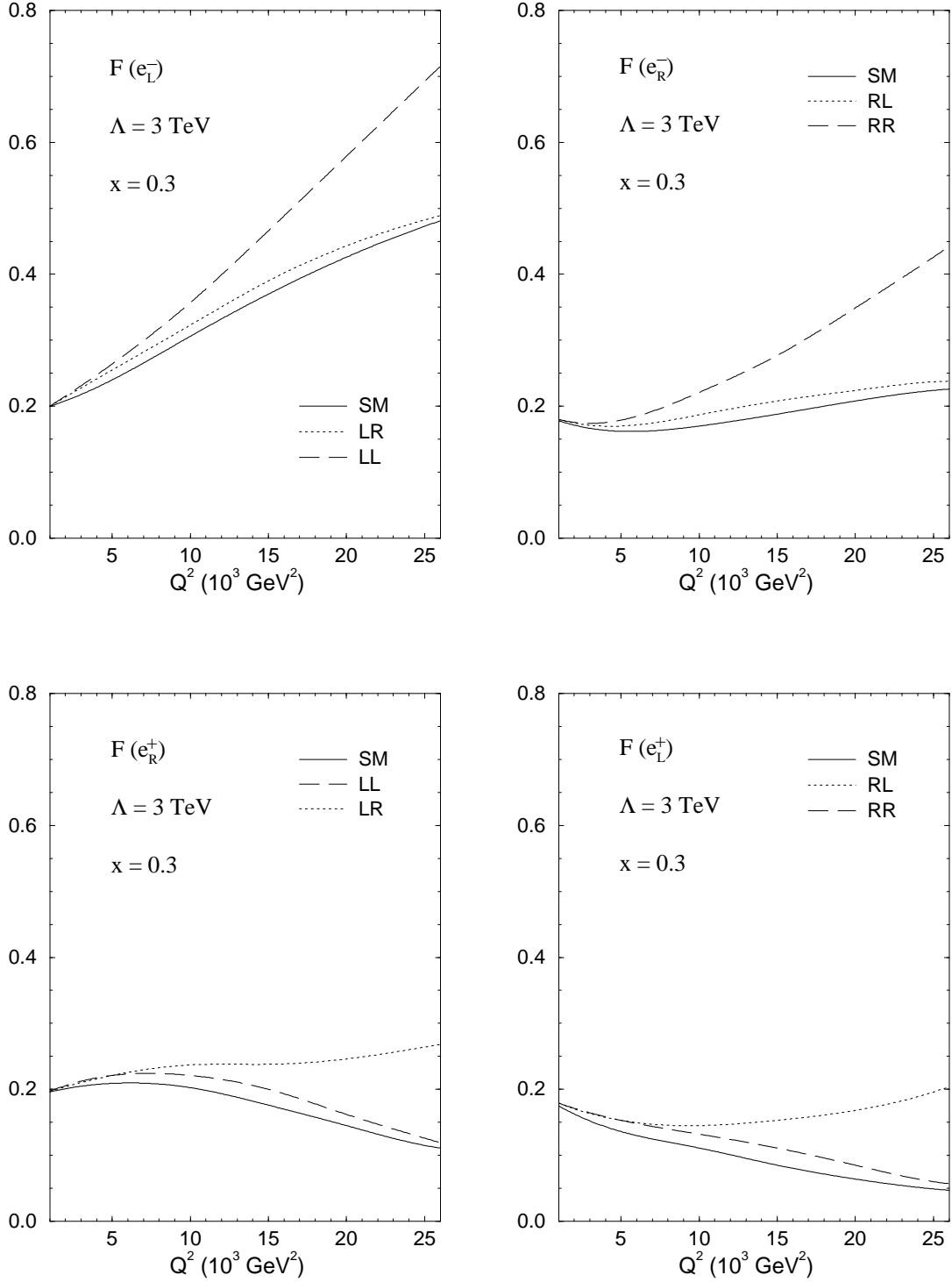}}
\end{picture}
\caption{Structure functions $F$ defined in the text, see Eq.\ (\ref{F}), for
polarized electron and positron beams. Notice that $LR$ and $RL$ contact terms
give a small contribution to the electron cross-section, but a large one to
the positron cross-section}
\end{figure}

\begin{table}
\begin{center}
\begin{tabular}{|c|c|c|}
\hline
\quad & $\Lambda^-$ ($TeV$) & $\Lambda^+$ ($TeV$)  \\
\hline
$LL$  &        $2.1$        &       $2.5$  \\
\hline
$RR$  &        $2.3$        &       $2.1$  \\
\hline
$LR$  &        $2.2$        &       $2.1$  \\
\hline
$RL$  &        $2.6$        &       $2.1$  \\
\hline
\end{tabular}
\end{center}
 \vspace*{0.3cm}
\caption{Lower bounds on the contact term mass scale obtained by the OPAL
collaboration. \cite{OPAL}}
\end{table}

There are, certainly, constrains on the values of $\Lambda$ from present
experiments. Recently OPAL has published bounds on $\Lambda$ from
measurements at LEP with a center of mass energy $\sqrt{s} = 130 - 136 \;
GeV$ and $161 \; GeV$ \cite{OPAL}. This bounds are shown in Table 1. Also
atomic parity violation measurements provide bounds on $\Lambda$. There, you
measure the weak charge given by:
\begin{equation}
\label{WC}
\begin{array}{ll}
Q_W = & -2 \left[
         \left( - \dis{1 \over 2} + \dis{4 \over 3} \sin^2 \theta_W \right)
         \left( 2Z + N \right) +
         \left( \dis{1 \over 2} - \dis{2 \over 3} \sin^2 \theta_W \right)
         \left( 2N + Z \right) \right] \\[0.3cm]
 \quad &  - \dis{1 \over \sqrt{2} G_F}
                                    \left[ \left( 2Z+N \right) \Delta \eta^{eu}
                                 + \left( 2N+Z \right) \Delta \eta^{ed}
\right],
\end{array}
\end{equation}
where $N$ and $Z$ are the number of neutrons and protons, respectively, in the
nucleus. The first term in the right hand side is the SM contribution and
\begin{equation}
\Delta \eta^{eq} = \eta^{eq}_{RR} - \eta^{eq}_{LL}
                 + \eta^{eq}_{RL} - \eta^{eq}_{LR} .
\end{equation}
The most stringent bounds on the mass scales can be obtained from the
$Q_W$ measurements in $^{133}_{55}Cs$:
\begin{equation}
Q_W = -71.04 \pm 1.81
\end{equation}
that should be compared with the SM prediction, $Q_W = -73.04$. Assuming
that only one of the operators in the lagrangian contributes, the bounds on
$\Lambda$ range between $7.4 \; TeV$ and $12.3 \; TeV$ \cite{DEANDREA}. These
values are too high to give a sensible modification of the
SM cross-section at HERA. However, these bounds can be easily avoided taking
the appropriate combinations of contact terms. Notice that this is not
fine tuning, but just determining the chiral structure of the new physics
couplings.

The low energy bounds on the mass scale still allows for values that clearly
are in better agreement with the experimental data than the SM predictions.
As an example we show in Fig.\ 3 the integrated cross-section as a function
of the minimum $Q^2$ for the Standard Model (solid line) and for an $eu$
contact term with $\Lambda_{LR} = \Lambda_{RL} = 3 \; TeV$, which avoids the
atomic parity violation bounds. Since only the $LR$ and $RL$ contact
terms are assumed to contribute to the cross-sections, the effects on the
electron cross-section are smaller than the ones in the positron
cross-section, in such a way that with the accumulated integrated luminosity
no observable deviation from the SM prediction is expected.
\begin{figure}[h]
\setlength{\unitlength}{1cm}
\begin{picture}(12.,10.27)
\epsfxsize=13cm
\put(1.64,-9.3){\epsfbox{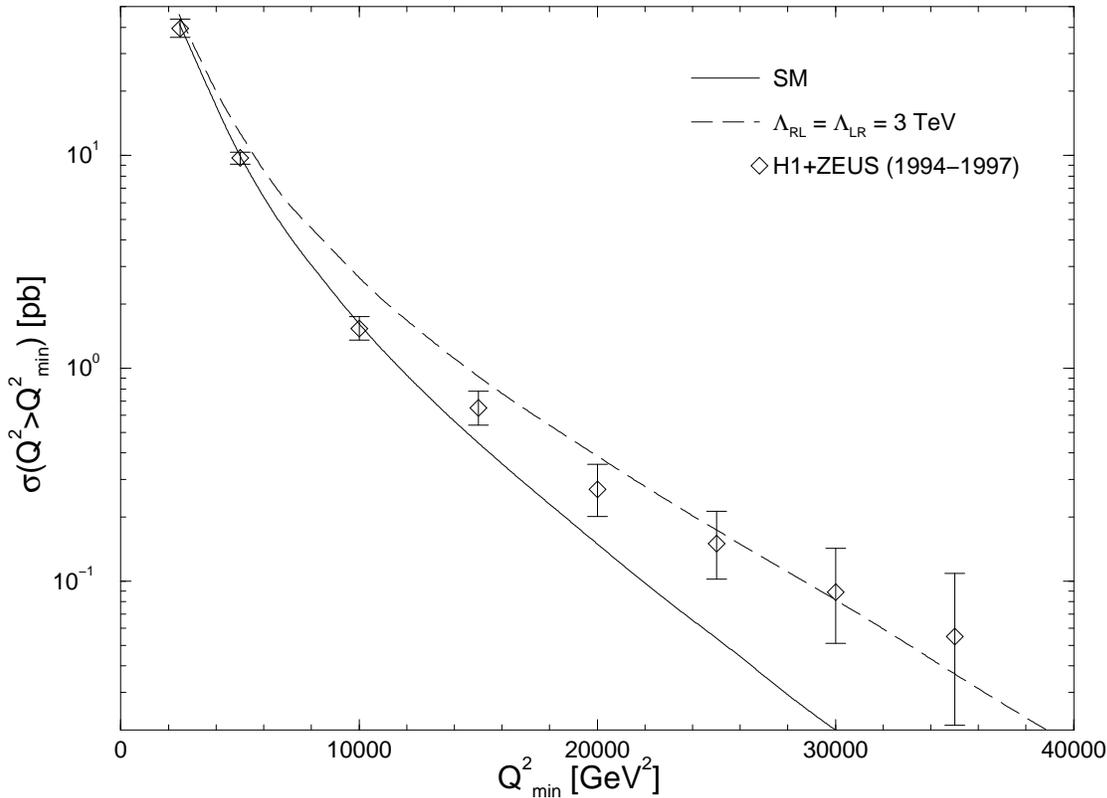}}
\end{picture}
\caption{Cross-section for $e^+ p \to e^+ X$ for $Q^2 \ge Q^2_{min}$ as
a function of $Q^2_{min}$ in the Standard Model (solid line) and with an
$eu$ contact term with $\Lambda_{RL} = \Lambda_{LR} = 3 \; TeV$. The combined
experimental data from H1 and ZEUS are also shown.}
\end{figure}

\subsection{Charged Currents}
The charged current cross-sections also receive contributions from contact
terms \cite{NOS}. The difference with the neutral current case is that
now only the
first term in Eq.\ (\ref{LAG}) contributes.

There are no direct experimental bounds on $\Lambda^{(3)}_{LL}$. The bounds
on $\Lambda_{LL}$ discussed in the previous section do not apply
directly because in neutral current processes one is actually measuring
the combination:
\begin{equation}
\Lambda_{LL} = \Lambda^{(3)}_{LL} + \Lambda^{(1)}_{LL} .
\end{equation}
It is, thus, important to measure both channels to disentangle the
contributions from both terms
in the lagrangian (\ref{LAG}). An indirect bound,
$\Lambda^{(3)}_{LL} > 10 \; TeV$ can be found from the unitarity of the
CKM matrix \cite{ALTARELLI}. This bound assumes that only the first family
is sensitive to contact terms.

The cross sections for $e^-p\to \nu X$ and $e^+p\to \bar{\nu} X$ can be
obtained
in a straightforward way :

\begin{equation}
\frac{d^2 \sigma_{e^-p\to \nu X}}{dQ^2 dx}
    =  \frac{1}{\pi} \left( \frac{G_F}{\sqrt{2}} \frac{M_W^2}{Q^2+M_W^2} -
   \frac{\pi}{2} \frac{\epsilon}{{\Lambda_{LL}^{(3)}}^2} \right)^2
 \sum_{i=1}^{2} \left[ u_i(x,Q^2)+
    (1-y)^2 \bar{d}_i(x,Q^2) \right]
     \label{electroncs}
\end{equation}
\begin{equation}
\frac{d^2 \sigma_{e^+p\to \bar{\nu} X}}{dQ^2 dx}
   = \frac{1}{\pi} \left( \frac{G_F}{\sqrt{2}} \frac{M_W^2}{Q^2+M_W^2} -
  \frac{\pi}{2} \frac{\epsilon}{{\Lambda_{LL}^{(3)}}^2}\right)^2
 \sum_{i=1}^{2} \left[ \bar{u}_i(x,Q^2) +
     (1-y)^2 d_i(x,Q^2) \right] \nonumber
     \label{positroncs}
\end{equation}

The contact term increases (decreases) the cross-sections for
$\epsilon = -1$ ($+1$) and the effects are larger for large values of
$Q^2$ and $x$.

We have performed a $\chi^2$ fit to the $Q^2$ distributions obtained by
H1 and ZEUS presented
by Elsen in Ref.\ \cite{CONF} \footnote{a similar fit to the electron and
positron data obtained before 1997 can be found in Ref.\ \cite{NOS}.}. The
fit to the H1 data does not improve with the inclusion of a contact term.
There is an excess of events at high $Q^2$ but at low $Q^2$ the SM prediction
lies above (although within one standard deviation) the measured
cross-section. The inclusion of a contact term can improve the agreement
between theory and experiment at high $Q^2$, but also produces a small increase
in the cross-section
at low $Q^2$ that spoils the quality of the fit. Thus from the H1 data
one can only obtain the following $95 \, \% C.L.$ bounds on the mass
scale:
\begin{equation}
\label{H1}
\begin{array}{ccc}
\Lambda_{LL}^{(3)+} \ge 3.5 \; TeV &
\Lambda_{LL}^{(3)-} \ge 3.4 \; TeV &
{\hbox{(From H1 1994-97 data).}}
\end{array}
\end{equation}

The situation is very different for the ZEUS data. Using the 1994-97 data
we now obtain a value
for $\eta_{LL}^{(3)}$ that is not compatible (within one standard deviation)
with zero, namely:
\begin{equation}
\label{ZEUSQ2}
\begin{array}{ccc}
\eta_{LL}^{(3)} = - 0.6 \pm 0.4 \; TeV^{-2} &
{\hbox{i.e.}} &
\Lambda_{LL}^{(3)-} = 4.6^{+3.3}_{-1.1} \; TeV
\end{array}
\end{equation}
We have also made a similar fit to the $x$ distribution measured by
ZEUS obtaining again a non-zero value for $\eta_{LL}^{(3)}$ compatible with
the previous one:
\begin{equation}
\label{ZEUSX}
\begin{array}{ccc}
\eta_{LL}^{(3)} = - 0.48^{+0.27}_{-0.33} \; TeV^{-2} &
{\hbox{i.e.}} &
\Lambda_{LL}^{(3)-} = 5.1^{+4.0}_{-1.0} \; TeV .
\end{array}
\end{equation}
These results have been obtained using the MRSA parton density parameterization
\cite{MRSA} and the mass of the $W$ boson has been fixed to the value given
by the Particle Data Book: $M_W = 80.33 \pm 0.15 \; GeV$ \cite{PDG}. We have
checked that the results of the fits do not change in an appreciable way when
moving the mass of the $W$ within its experimental error.

\section{Summary}
There has been a lot of work during the last years on the capabilities of
HERA to search for physics beyond the Standard Model. In particular,
after the publication of the excess of events at high $Q^2$ observed
by the two experiments, H1 and ZEUS, the number of papers trying to
explain them in terms of new physics has exploded. Since we cannot cover
everything in a single talk we have concentrated on two subjects: search
for the Minimal Supersymmetric Standard Model with $R$-parity conservation
and the effects of contact terms in neutral and charged current processes.
Other interpretations of the excess of events in terms of leptoquark
production and squark production in $R$-parity breaking supersymmetry
have been discussed in other talks at this school
\cite{SPIESBERGER,KALINOWSKI}.

The most promising process to search for $R$-parity conserving SUSY is
the associate production of a selectron and an squark, with these
sfermions decaying into the corresponding fermions and the lightest
neutralino. The experimental signature would be a final state containing
an electron, a jet and missing energy. We have shown the exclusion bounds
that can be obtained at HERA assuming an integrated
luminosity of $100$ and $500 \; pb^{-1}$. These bounds turn out to be
competitive and complementary to the ones that can be obtained at
LEP and TEVATRON.

It is not clear for the moment whether the excess of events observed at
HERA in neutral and charged current processes is really a signal of new
physics or just a statistical fluctuation. In case it is a signal for new
physics it is difficult to explain it in terms of the production of a single
resonance. Instead it can be explained in terms of contact terms, i.e.
dimension $6$ four-fermion operators. The contact terms required to improve
the agreement between the theoretical predictions and the observed number
of events with respect to the SM predictions are
compatible with low energy constraints, such as atomic parity violation
measurements, and high energy bounds from LEP and TEVATRON.

One of us, F.C., thanks K. Kolodziej , J. Sladkowski and M. Zraleck for their
hospitality during this enjoyable school. This work was partially
supported by CICYT under contract AEN96-1672 and by Junta de Andalucia.

\end{document}